\documentclass[12pt]{article}
\usepackage[]{graphicx,amssymb,array}

\begin{document}

   \begin{center}
       {\Large Analytic parametrizations of the non-perturbative Pomeron 
              and QCD-inspired models\footnote{Talk given by B. Nicolescu at the 9th 
    International Conference on Elastic and Diffractive 
            Scattering (9th Blois Workshop), Pruhonice, Czech Republic, 9-15 June 2001}}
	      
	      \vspace{1cm}
         B. Nicolescu$^a$,  
          J. R. Cudell$^b$, 
        V. V. Ezhela$^c$,
       P. Gauron$^a$,\\
       K. Kang$^d$, 
       Yu.~V.~Kuyanov$^c$,
        S.~B. Lugovsky$^c$,\\
	and N. P. Tkachenko$^c$
   \end{center}
   
     \noindent $^a$ LPNHE\footnote{Unit\'e de Recherche des Universit\'es Paris 6 et Paris 7, 
    Associ\'e au CNRS}, Universit\'e Pierre et Marie Curie, Tour 12 E3, 4 
    Place Jussieu, 75252 Paris Cedex 05, France
    
    \noindent $^b$ Institut de Physique, B\^at. B5, Universit\'e de 
    Li\`ege, Sart Tilman, B4000 Li\`ege, Belgium
    
    \noindent $^c$ COMPAS group, IHEP, Protvino, Russia
       
    \noindent $^d$ Physics Department, Brown University, Providence, RI, 
    U.S.A.

    \begin{center}
        (COMPETE Collaboration)
    \end{center}

    \begin{abstract}
        We consider several classes of analytic parametrizations of hadronic 
        scattering amplitudes, and compare their predictions to all available 
        forward data ($pp, \bar pp, \pi p, Kp,$ 
 $\gamma p, \gamma\gamma,\Sigma 
        p$). Although these parametrizations are very close for $\sqrt{s}\ge 
        9$ GeV, it turns out that they differ markedly at low energy, where 
        a universal Pomeron term $\sim\ln^2s$ enables one to extend the fit 
        down to $\sqrt{s}=4$ GeV.
    \end{abstract}

\section{The COMPETE project}
Analytic parametrizations of forward $(t=0)$ hadron scattering 
amplitudes is a well-established domain in strong interactions. 
The basic idea is to implement as much as possible general 
principles - analyticity, unitarity, crossing -symmetry and positivity 
of total cross-sections  - supplemented by some other general 
properties like the connection between Regge poles and resonance 
masses. 
In the absence of an explicit form of the forward amplitude derived 
from QCD, such an approach could provide 
a useful tool for studying non-perturbative physics.

However, in the past, the phenomenology of forward scattering had 
quite a high degree of arbitrariness.

Some of the arbitrariness in question are :
\begin{itemize}
    \item  [i)]
    An excessive focus on $pp$ and $\bar pp$ scattering. When other 
    reactions were studied they were often analyzed one by one.
    Of course, the extraction of the free parameters and their 
    physical interpretation lead to contradictory conclusions.

    \item  [ii)]
    Important physical constraints are often mixed with less general 
    or even ad-hoc properties.

    \item  [iii)]
    The cut-off in energy, defining the region of applicability of the 
    high-energy models, differs from one author to the other.

    \item  [iv)]
    The set of data considered by different authors is sometimes not 
    the same, even if the cut-off in energy is the same : arbitrary 
    exclusions of experimental data are performed.

    \item  [v)]
    No rigorous connection is made between the number of parameters 
    and the number of data points.
    The requirement of the smallest possible number of free parameters 
    does not necessarily signify the physical validity of the model.

    \item  [vi)]
    No attention was paid to the necessity of the stability of 
    parameter values when different blocks of data or different sets 
    of observables were used.

    \item  [vii)]
    The experiments were performed in the past in quite a chaotic 
    way : huge gaps are sometimes present between low-energy and 
    high-energy domains or inside the high-energy domain itself.
    As the data at low energies are often quite precise we are in the 
    paradoxical situation of jumping to conclusions concerning the 
    high-energy structure of scattering amplitudes on the basis of 
    information concentrated at low energies.
\end{itemize}

This is only a partial list and is of course not exhaustive.

The COMPETE (\underline{CO}mputerized \underline{M}odels and 
\underline{P}arameter \underline{E}valuation for \underline{T}heo\-ry 
and \underline{E}xperiment) project \cite{Cudell,Kang} tries to cure 
as much as possible the above discussed arbitrariness.

First of all, we include in our analysis \cite{Cudell} all the 
existing forward data ($pp, \bar pp,$\\ 
$\pi p, Kp, \gamma p, 
\gamma\gamma,\Sigma p$) for the total cross-sections $\sigma(s)$ and 
the $\rho$-parameter
\begin{equation}
    \rho(s)=\left.\frac{ReF(s,t)}{ImF(s,t)}\right\vert_{t=0}
    \label{eq:1}
\end{equation}
by developing the procedure initiated in Refs. \cite{Cudell00} and 
\cite{Cudell97}, thus taking problem i) into account.

The problem ii) is cured by studying a large variety of variants of a 
given model, each variant corresponding to a given set of physical 
properties.

The problems iii)-vi) are treated by defining appropriate numerical 
indicators \cite{Cudell,Kang}. The $\chi^2/dof$ criterium is not able, 
by itself, to cure the difficulties iii)-vi): new indicators have 
to be defined.

Once these indicators are defined, an appropriate sum of their 
numerical values is proposed in order to establish the \textbf{rank} 
of the model under study : the highest the numerical value of the rank 
the better the model under consideration.

The problem vii) is not yet solved: theory or phenomenology can not 
replace the experiment itself. We very much hope that, in the future, 
experiments (e.g. at RHIC) will scan the available region of energies 
in small energy steps.

The final aim of the COMPETE project is to provide our community with a 
periodic cross assessments of data and models via computer-readable 
files on the Web \cite{web}.

\section{The form of the forward scattering amplitudes}
We consider the following exemplar cases of the imaginary part of the 
scattering amplitudes :
\begin{equation}
    ImF^{ab}=s\sigma_{ab}(s)=P_{1}^{ab}(s)+P_{2}^{ab}(s)+R_{+}^{ab}(s)
    \pm R_{-}^{ab}(s)
    \label{eq:1prime}
\end{equation}
where : \\
- the $\pm$ sign in formula \ref{eq:1prime} corresponds to 
antiparticle(resp. particle) - particle scattering amplitude.\\
- $R_{\pm}$ signify the effective secondary-Reggeon 
($(f,a_{2}),\ (\rho,\omega)$) contributions to the 
even(odd)-under-crossing amplitude
\begin{equation}
    R_{\pm}(s)=Y_{\pm}\left(\frac{s}{s_{1}}\right)^{\alpha_{\pm}}
    \label{eq:2}
\end{equation}
$Y$ being a constant residue, $\alpha$ - the reggeon intercept and 
$s_{1}$ - a scale factor fixed at 1 GeV$^2$ ;\\
- $P_{1}(s)$ is the contribution of the Pomeron Regge pole
\begin{equation}    
    P_{1}^{ab}(s)=C_{1}^{ab}\left(\frac{s}{s_{1}}\right)^{\alpha_{P_{1}}},
    \label{eq:3}
\end{equation}
$\alpha_{P_{1}}$ is the Pomeron intercept
$
    \alpha_{P_{1}}=1,
$
and $C^{ab}$ are constant residues.\\
- $P_{2}^{ab}(s)$ is the second component of the Pomeron 
corresponding to three different $J$-plane singularities :
\begin{itemize}
    \item  [a)] a Regge simple - pole contribution
    \begin{equation}    
    P_{2}^{ab}(s)=C_{2}^{ab}\left(\frac{s}{s_{1}}\right)^{\alpha_{P_{2}}}, 
    \ \mbox{with}\ 
    \alpha_{P_{2}}=1+\epsilon,\  \epsilon >0, \ 
    \mbox{and }C_{2}^{ab}\mbox{ const. ;} 
        \label{eq:5}
    \end{equation}

    \item  [b)] a Regge double-pole contribution
    \begin{equation}                        
        P_{2}^{ab}(s)=s\left[A^{ab}+B^{ab}\ln\left(\frac{s}{s_{1}}
        \right)\right], 
        \mbox{with }A^{ab}\mbox{ and }B^{ab}\mbox{ const. ;}
        \label{eq:7}
    \end{equation}

    \item  [c)] a Regge triple-pole contribution
    \begin{equation}
        P_{2}^{ab}(s)=s\left[A^{ab}+B^{ab}\ln^2\left(\frac{s}{s_{0}}
        \right)\right],
        \label{eq:8}
    \end{equation}
where $A^{ab}$ and $B^{ab}$ are constants and $s_{0}$ is an arbitrary 
scale factor.
\end{itemize}

The real part of the $F^{ab}$ amplitude is obtained from formula
\ref{eq:1prime} via the well-known substitution rule
$
    s\to se^{-i\pi/2}
$
or, in an equivalent way, from the derivative relations \cite{Kang75}.

In other words, (\ref{eq:1prime}) with (\ref{eq:3}),(\ref{eq:5}) and 
(\ref{eq:8}) represents three 
exemplar cases for the asymptotic behaviour of total cross-sections :
\begin{equation}
    \sigma\mathop{\longrightarrow}_{s\to\infty} \mbox{const.},\ \ln s,\ 
    \ln^2s,
    \label{eq:10}
\end{equation}
while (\ref{eq:1prime}) with (\ref{eq:5}) is the special case
\begin{equation}
    \sigma\mathop{\longrightarrow}_{s\to\infty} s^{1+\epsilon}
    \label{eq:11}
\end{equation}
corresponding to the violation of the Froissart 
axiomatic bound \cite{Heisen52,Frois61,Martin66}.
It however proved to be useful for studying the data at 
non-asymptotic energies \cite{Land92} and therefore we will include 
it as well, and assume that unitarity corrections will 
restore this bound.
In fact, the three exemplar cases (\ref{eq:3}), (\ref{eq:7}) and 
(\ref{eq:8}) are 
representatives of an infinite class of analytic parametrizations 
leading to a $(\ln s)^{\beta_{+}}\ (0\leq\beta_{+}\leq 2)$ behaviour 
of the total cross-sections \cite{Kang75}, so our study is quite general.

As can be seen from formulae (\ref{eq:1prime})-(\ref{eq:8}), we 
consider that the Pomeron has two soft components , a property in 
agreement with perturbative QCD : recently, Bartels, Lipatov and Vacca 
\cite{Bartels00} discovered that there are, in fact, two types of 
Pomeron in LLA : besides the well-known BFKL pomeron associated with 
2-gluon exchanges, and with an intercept bigger than 1, there
is a second one associated with $C=+1$ three-gluon exchanges and 
having an intercept precisely located at 1. The case 
(\ref{eq:3})-(\ref{eq:5}) is directly inspired by this result, while 
the cases (\ref{eq:3}) with (\ref{eq:7}) and (\ref{eq:8}) are subtler 
because both the components of the Pomeron satisfy the unitarity 
constraint via Regge singularities all located at $J=1$ 
\cite{Gauron00,BN01}.

\section{Results of the fits}
We consider all the existing forward data for $pp,\bar pp, \pi p, Kp, 
\gamma\gamma$ and $\Sigma p$ scatterings \cite{web}.
The number of data points is : 904, 742, 648,  569, 498, 453, 397, 
329 when the cut-off in energy is 3, 4, 5, 6, 7, 8, 9, 10 GeV 
respectively.
When only $\sigma$ data are considered the number of points is 
reduced to 726, 581, 507, 434, 369, 331, 285, 230 respectively.

A large number of variants were studied and classified \cite{Cudell}. 
All definitions and numerical details can be found in Ref.\cite{Cudell}.

In order to describe our main results let us first introduce some 
notations.

The\hfill 2-component\hfill Pomeron\hfill classes\hfill of\hfill models\hfill
are\hfill
    RRPE,\hfill RRPL\hfill and \\ RRPL2,
where by RR we denote the two effective secondary-reggeon 
contributions, by P - the contribution of the Pomeron Regge-pole 
located at $J=1$, by E - the contribution of the Pomeron Regge-pole 
located at $J=1+\epsilon$, by L - the contribution of the component of 
the Pomeron, located at $J=1$ (double pole) and leading to a $\ln s$ 
behaviour of $\sigma$, and by L2 - the contribution of the 
component of the Pomeron located at $J=1$ (triple pole) and leading to 
a $\ln^2s$ behaviour of $\sigma$.

We also studied the 1-component Pomeron classes of models
    RRE, RRL and RRL2.

Our two main conclusions are the following :
\begin{itemize}
    \item  [1.]
    The familiar RRE model is rejected at the 98\% C.L. when models 
    which achieve a $\chi^2/dof$ less than 1 for $\sqrt{s}\geq 5$ GeV 
    are considered.

    \item  [2.]
    The best fits are given by models that contain a triple pole at 
    $J=1$, which then produce $\log^2s$, $\log s$ and constant terms 
    in the total cross section.
\end{itemize}510

In his seminal paper, published in 1952 \cite{Heisen52}, Heisenberg 
was the first to introduce the $\ln^2s$ dependence of the hadronic 
total cross-sections, by simple and familiar arguments. 
His main assumption was that the fraction of energy which goes into 
the meson field (whose lower limit is given by the minimal energy for the 
creation of at least a pair of pions) is proportional to the overlap of 
the meson fields in the nucleon.
Interestingly enough, the Heisenberg result is a \textbf{finite-energy} 
one : his total cross-section is a quadratic form in $\ln s$. Only a 
decade later, a rigorous proof of the Heisenberg result, but only in 
the case of asymptotic energies, was given \cite{Frois61,Martin66}.

A striking result, directly seen from the data, is the 
\textbf{universality} of the $\ln^2(s/s_{0})$ terms in eq. \ref{eq:8} : 
$B^{ab}$ and $s_{0}^{ab}$ are independent of the hadrons involved in 
the scattering \cite{Gauron00,BN01}. 
A general theoretical proof of this property does not exist yet. 
It is tempting to speculate that, after unitarization is performed 
in the gluon sector, the BFKL Pomeron would finally lead to a universal 
Heisenberg-type Pomeron, exclusively connected with the gluon sector.

As a typical example of high-rank RRPL2 models let us consider in 
more detail the RRPL2$_{u}$ (21) model (where the $u$ index denotes the 
above discussed universality property and the number between 
parenthesis denotes the number of the free parameters.

The explicit forms of the observables are the following :

\begin{eqnarray}
    \sigma_{pp} & = & Z_{pp}+B\ln^2 \left(\frac{s}{s_{0}}\right)+
    Y_{+}^{pp}s^{-\eta_{+}}-Y_{-}^{pp}s^{-\eta_{-}},\\
    \sigma_{\bar pp} & = & Z_{pp}+B\ln^2 \left(\frac{s}{s_{0}}\right)+
    Y_{+}^{pp}s^{-\eta_{+}}+Y_{2-}^{pp}s^{-\eta_{-}},  \\
    \sigma_{\pi^+p} & = & Z_{\pi p}+B\ln^2 \left(\frac{s}{s_{0}}\right)+
    Y_{+}^{\pi p}s^{-\eta_{+}}-Y_{-}^{\pi p}s^{-\eta_{-}}, \\
    \sigma_{\pi^-p} & = & Z_{\pi p}+B\ln^2 \left(\frac{s}{s_{0}}\right)+
    Y_{+}^{\pi p}s^{-\eta_{+}}+Y_{-}^{\pi p}s^{-\eta_{-}},   \\
    \sigma_{K^+p} & = & Z_{Kp}+B\ln^2 \left(\frac{s}{s_{0}}\right)+
    Y_{+}^{Kp}s^{-\eta_{+}}-Y_{-}^{Kp}s^{-\eta_{-}},  \\
   \sigma_{K^-p} & = & Z_{Kp}+B\ln^2 \left(\frac{s}{s_{0}}\right)+
    Y_{+}^{Kp}s^{-\eta_{+}}+Y_{-}^{Kp}s^{-\eta_{-}},   \\
    \sigma_{\gamma p} & = & Z_{\gamma p}+\delta B\ln^2 \left(\frac{s}{s_{0}}\right)+
    Y_{+}^{\gamma p}s^{-\eta_{+}}, \\
    \sigma_{\gamma \gamma} & = & Z_{\gamma \gamma}+\delta^2B\ln^2 
    \left(\frac{s}{s_{0}}\right)+
    Y_{+}^{\gamma \gamma}s^{-\eta_{+}},    \\
    \sigma_{\Sigma^-p} & = & Z_{\Sigma p}+B\ln^2 \left(\frac{s}{s_{0}}\right)+
    Y_{+}^{\Sigma p}s^{-\eta_{+}}-Y_{-}^{\Sigma p}s^{-\eta_{-}}, \\
    \rho_{pp}\sigma_{pp} & = & \pi B\ln\left(\frac{s}{s_{0}}\right)                
    -\frac{Y_{+}^{pp}s^{-\eta_{+}}}
    {\tan\left[\frac{1-\eta_{+}}{2}\pi\right]}
    -\frac{Y_{-}^{pp}s^{-\eta_{-}}}
    {\cot\left[\frac{1-\eta_{-}}{2}\pi\right]},
    \label{eq:21}  \\
     \rho_{\bar pp}\sigma_{\bar pp} & = & \pi B\ln\left(\frac{s}{s_{0}}\right)                
    -\frac{Y_{+}^{pp}s^{-\eta_{+}}}
    {\tan\left[\frac{1-\eta_{+}}{2}\pi\right]}
    +\frac{Y_{-}^{pp}s^{-\eta_{-}}}
    {\cot\left[\frac{1-\eta_{-}}{2}\pi\right]},
    \label{eq:22}  \\
    \rho_{\pi^+p}\sigma_{\pi^+p} & = & \pi B\ln\left(\frac{s}{s_{0}}\right)                
    -\frac{Y_{+}^{\pi p}s^{-\eta_{+}}}
    {\tan\left[\frac{1-\eta_{+}}{2}\pi\right]}
    -\frac{Y_{-}^{\pi p}s^{-\eta_{-}}}
    {\cot\left[\frac{1-\eta_{-}}{2}\pi\right]},
    \label{eq:23}  \\
     \rho_{\pi^-p}\sigma_{\pi^-p} & = & \pi B\ln\left(\frac{s}{s_{0}}\right)                
    -\frac{Y_{+}^{\pi p}s^{-\eta_{+}}}
    {\tan\left[\frac{1-\eta_{+}}{2}\pi\right]}
    +\frac{Y_{-}^{\pi p}s^{-\eta_{-}}}
    {\cot\left[\frac{1-\eta_{-}}{2}\pi\right]},
    \label{eq:24}  \\
    \rho_{K^+p}\sigma_{K^+p} & = & \pi B\ln\left(\frac{s}{s_{0}}\right)                
    -\frac{Y_{+}^{Kp}s^{-\eta_{+}}}
    {\tan\left[\frac{1-\eta_{+}}{2}\pi\right]}
    -\frac{Y_{-}^{Kp}s^{-\eta_{-}}}
    {\cot\left[\frac{1-\eta_{-}}{2}\pi\right]},
    \label{eq:25}  \\
    \rho_{K^-p}\sigma_{K^-p} & = & \pi B\ln\left(\frac{s}{s_{0}}\right)                
    -\frac{Y_{+}^{Kp}s^{-\eta_{+}}}
    {\tan\left[\frac{1-\eta_{+}}{2}\pi\right]}
    +\frac{Y_{-}^{Kp}s^{-\eta_{-}}}
    {\cot\left[\frac{1-\eta_{-}}{2}\pi\right]},
    \label{eq:27}
\end{eqnarray}

\begin{equation}
  \mbox{where }\quad \eta_{\pm}=1-\alpha_{\pm}
   \mbox{  and }
   Z_{ab}\equiv C_{1}^{ab}+A^{ab}.
    \label{eq:28}
\end{equation}

The numerical values of the free parameters are given in Table 1.

This model gives a very good $\chi^2/dof=0.973$ and has a high rank 
point with a total indicator value $P^M=222$. 
It describes well the data for $\sqrt{s}\geq 5$ GeV (648 
data points), respecting all our numerical criteria for the selection 
of acceptable models.

In order to understand the content of this empirically-found 
universality property let us consider the most general form of the 
two - component PL2 Pomeron as given by (\ref{eq:3}) and (\ref{eq:8}) 
with $Z_{ab}$ defined by (\ref{eq:28}) :
\begin{equation}
    \frac{1}{s}\left(P_{1}^{ab}+P_{2}^{ab}\right)=Z_{ab}+B^{ab}
    \ln^2\left(
    \frac{s}{s_{0}^{ab}}
    \right).
    \label{eq:29}
\end{equation}
This form contains 18 free parameters :\\
$Z_{pp}, Z_{\pi p}, Z_{Kp}, Z_{\Sigma p}, Z_{\gamma p}, 
Z_{\gamma\gamma}$ ;
$B^{pp}, B^{\pi p}, B^{Kp}, B^{\Sigma p}, B^{\gamma p}, 
B^{\gamma\gamma}$ ;
$s_{0}^{pp}, s_{0}^{\pi p}, s_{0}^{Kp}, s_{0}^{\Sigma p},\\ s_{0}^{\gamma p}, 
s_{0}^{\gamma\gamma}$.

The content of our empirical universality property is threefold :
\begin{itemize}
    \item  [1)] a \textbf{finite energy} property : all scale factors $s_{0}$ 
    are the same :
    \begin{equation}
        s_{0}\equiv 
        s_{0}^{pp}=s_{0}^{\pi p}=s_{0}^{Kp}=s_{0}^{\Sigma 
        p}=s_{0}^{\gamma p}=s_{0}^{\gamma\gamma} ;
        \label{eq:30}
    \end{equation}

    \item  [2)] an \textbf{asymptotic} property : all hadron-hadron 
    cross-sections are the same at infinite energies :
    \begin{equation}
        \sigma_{ab}\mathop{\longrightarrow}_{s\to\infty}B\cdot\ln^2s,
        \label{eq:31}
    \end{equation}
    where
    \begin{equation}
        B\equiv B^{pp}=B^{\pi p}=B^{Kp}=B^{\Sigma p} ;
        \label{eq:32}
    \end{equation}

    \item  [3)] a second (factorization) \textbf{asymptotic} property :
    \begin{equation}
        \sigma_{\gamma p}\mathop{\longrightarrow}_{s\to\infty}
        \sqrt{\sigma_{\gamma\gamma}}\cdot\sqrt{\sigma_{pp}}
        =\delta B\ln^2s
        \label{eq:33}
    \end{equation}
\end{itemize}
The nine constraints resulting from (\ref{eq:30})-(\ref{eq:33}) reduce the 
number of free parameters of the PL2 Pomeron from 18 to 9.

The secondary-reggeon contributions involve 12 more parameters :
    $Y_{\pm}^{pp}$, $Y_{\pm}^{\pi p}, Y_{\pm}^{Kp},Y_{\pm}^{\Sigma p},
    Y_{+}^{\gamma p},Y_{+}^{\gamma \gamma};\ \alpha_{+},\alpha_{-}$.

Hence we have a total of 21 parameters for the RRPL2$_{u}$ (21) model. It is 
interesting to note that the high-energy (Pomeron) part is more 
constrained than the low-energy (secondary-reggeon) part.
At the RHIC, LHC and cosmic-ray energies the RRPL2$_{u}$ model has 
only 9 free parameters.
The cosmic-ray experimental data (not included in fits) are very well 
described by our high-rank model.

\begin{center}
    \begin{tabular}{|c|c|c|}
        \hline
        Model & \multicolumn{2}{|c|}{RRPL2$_{u}$ }  \\
        \hline
        $\chi^2/dof$ & \multicolumn{2}{|c|} {0.973}    \\
        \hline
        CL[\%] &  \multicolumn{2}{|c|} {67.98}   \\
        \hline
        Parameter & Mean & Uncertainty  \\
        \hline
        $s_{0}$ & 34.0 & 5.4  \\
        B & 0.3152 & 0.0095  \\
        $\alpha_{+}$ & 0.533 & 0.015  \\
        $\alpha_{-}$ & 0.4602 & 0.0064  \\
        $Z_{pp}$ & 35.83 & 0.40  \\
        $Z_{\pi p}$ & 21.23 & 0.33  \\
        $Z_{Kp}$ & 18.23 & 0.30  \\
        $Z_{\Sigma p}$ & 35.6 & 1.4  \\
        $Z_{\gamma p}$ & 0.109 & 0.021  \\
        $Z_{\gamma\gamma}$ & 0.075 & 0.026  \\
        $Y_{+}^{pp}$ & 42.1 & 1.3  \\
        $Y_{-}^{pp}$ & 32.19 & 0.94  \\
        $Y_{+}^{\pi p}$ & 17.8 & 1.1  \\
        $Y_{-}^{\pi p}$ & 5.72 & 0.16  \\
        $Y_{+}^{Kp}$ & 5.72 & 1.40  \\
        $Y_{-}^{Kp}$ & 13.13 & 0.38  \\
        $Y_{+}^{\Sigma p}$ & -250. & 130.  \\
        $Y_{-}^{\Sigma p}$ & -320. & 150.  \\
        $Y_{+}^{\gamma p}$ & 0.0339 & 0.0079  \\
        $Y_{+}^{\gamma \gamma}$ & 0.00028 & 0.00015  \\
        $\delta$ & 0.00371 & 0.00035 \\
        \hline
    \end{tabular}    
\end{center}
\begin{center}
    Table 1. The numerical values of the free parameters, 
    $\chi^2/dof$ and the confidence level CL in the case of the
    RRP2$_{u}$ (21) model ($\sqrt{s}\geq 5$ GeV).
\end{center}   


\begin{center}
    \begin{tabular}{|c|c|c|c|c|}
        \hline
         & {\small Region} of & {\small Numerical value} & 
	 {\small $\alpha_{+}-\alpha_{-}$} & 
         {\small Remarks}  \\
         & {\small validity} & {\small of the rank points} &  &   \\
        \hline
        {\small RRPL2$_{u}$(21)} & {\small $\sqrt{s}\geq 5$ GeV} & 222 & $\simeq$ 
        0.07 & -  \\
        \hline
        {\small RRPL(21)} & {\small $\sqrt{s}\geq 5$ GeV} & 173 & $\simeq$ 0.33  & 
        {\small $Z_{pp},Z_{\pi p},Z_{Kp}<0$}\\
        \hline
        {\small RRPE$_{u}$(19)} & {\small $\sqrt{s}\geq 8$ GeV} & 158 & $\simeq$ 0.20 & 
        {\small $Z_{\pi p},Z_{Kp}<0$}\\
        \hline
    \end{tabular}
\end{center}

\begin{center}
    Table 2. Comparison between 3 high-rank representative models.
\end{center}

\newpage

In Table 2 we compare three representative models for each class of 
2-compo\-nent Pomeron : RRPL2$_{u}$ (21), RRPL (21) and RRPE$_{u}$ (19).

It is seen from Table 2 that the RRPL2$_{u}$ (21) model respects the 
Regge exchange-degeneracy\hfill to\hfill a\hfill very\hfill good\hfill
approximation,\hfill in\hfill contrast\hfill with\hfill the\\ 
RRPE$_{u}$~(19) and RRPL (21) models. 
The latter model leads to a huge violation, incompatible with the 
masses of the resonances on the respective Regge trajectories. 
This difficulty of the RRPE$_{u}$ 
(19) and RRPL (21) models is algebraically correlated with the 
negativity of the $Z$ parameters (see Table 1), a suspect feature 
which could lead - via factorization of Regge residues - to negative 
cross-sections in the case of the scattering involving $s,\ c,\ b$ 
and $t$ quarks.

This point reinforces our conclusion that the solution RRPL2$_{u}$ (21) 
is the best one, currently meeting all theoretical, phenomenological 
and numerical requirements.

\section{Future prospects}\label{sec:}
One problem remaining in the analysis of the forward data is the 
difficulty in adequately fitting the data for the $\rho$ parameter in 
$pp$ and in $\pi^+p$ reactions. 
The extraction of the $\rho$ data from the measurements of the 
differential cross sections data at small $t$ is a delicate problem. 
A reanalysis of these data may be needed, but it will call for simultaneous 
fits to the total cross-section data and to the elastic differential 
cross-sections in the Coulomb-nuclear interference region and in the diffractive 
cones, hence an extension of the parametrization considered here to the 
non-forward region.

\end{document}